\documentclass[prb,endfloats,eqsecnum]{revtex4}
\usepackage{graphicx}

\newcommand{\nbb}{{\mbox{\boldmath{$\nabla$}}}}
\newcommand{\bp}{{\bf p}}
\newcommand{\bq}{{\bf q}}
\newcommand{\br}{{\bf r}}

\newcommand{\bff}{{\bf F}}

\newcommand{\ep}{\varepsilon}

\newcommand{\e}{{\rm e}}

\newcommand{\nn}{\nonumber}
\newcommand{\la}{\label} 

\newcommand{\be}{\begin{equation}}
\newcommand{\ee}{\end{equation}}
\newcommand{\ba}{\begin{eqnarray}}
\newcommand{\ea}{\end{eqnarray}}

\begin{document}
\title{Forward and non-forward symplectic 
integrators\\ in solving classical dynamics problems}

\author{Siu A. Chin}

\affiliation{Department of Physics, Texas A\&M University,
College Station, TX 77843, USA}

\begin{abstract}
Forward time step integrators are splitting algorithms with
only positive splitting coefficients. When used in solving physical evolution
equations, these positive coefficients correspond to positive
time steps. Forward algorithms are essential for solving 
time-irreversible equations that cannot be evolved using
backward time steps. However, forward integrators 
are also better in solving time-reversible equations of 
classical dynamics by tracking as closely as possible
the physical trajectory. This work compares in detail various 
forward and non-forward fourth-order integrators using three, fourth,
five and six force evaluations. In the case of solving the 2D Kepler 
orbit, all non-forward integrators are optimized by simply minimizing 
the size of their backward time steps

\end{abstract}
\maketitle

\section {Introduction}
Many physical evolution equations are of the form
\be
{{\partial w}\over{\partial t}}=(T+V)w,
\la{gen}
\ee
where $T$ and $V$ are noncommuting operators.
Important examples include the imaginary time Schr\"odinger equation
\be
{{\partial\psi}\over{\partial \tau}}=
({{1}\over{2}}\nabla^2-V)\psi
\la{imeq}
\ee
and the Fokker-Planck equation
\be
{\partial\over{\partial t}} \rho({\bf x},t)
={1\over 2}\nabla^2 \rho({\bf x},t)
- \nabla\cdot[ {\bf v}({\bf x})\rho({\bf x},t)].
\label{fpeq}
\ee
Because the diffusion kernel $\nabla^2$ cannot be evolved 
backward in time, both of these are {\it time-irreversible} evolution equations. 
Aside from
these obvious examples of (\ref{gen}), any pair of equations of the form
\be
\frac{d\bq}{dt}={\bf v}(\bp),\qquad \frac{d\bp}{dt}={\bf F}(\bq),
\la{eqpair}
\ee
can also be casted into the form (\ref{gen}). This is because
the evolution of a general function $W(\bp,\bq)$ (including
$\bq$ and $\bp$ themselves) can
be formulated as
\ba
\frac{dW}{dt}&=&\frac{\partial W}{\partial \bq}\cdot\frac{d\bq}{dt}
+\frac{\partial W}{\partial \bp}\cdot\frac{d\bp}{dt}\nn\\
&=&\Bigl({\bf v}(\bp)\cdot\frac{\partial}{\partial \bq}
+{\bf F}(\bq)\cdot\frac{\partial}{\partial \bp}\Bigr)W,
\la{evol}
\ea
from which one can identify
\be
T={\bf v}(\bp)\cdot \frac{\partial}{\partial \bq}
\quad{\rm and}\quad
V
={\bf F}(\bq)\cdot\frac{\partial}{\partial \bp}.
\la{tvop}
\ee
Classical Hamiltonian dynamics corresponds to
\be
{\bf v}(\bp)=\frac\bp{m}
\ee
and the resulting evolution (\ref{eqpair})
is {\it time-reversible}. 
 
The generic evolution equation (\ref{gen}) can be solved iteratively 
\be
w(t+\ep)={\rm e}^{\ep(T+V)} w(t)
\ee
by approximating the short time evolution operator  
${\rm e}^{\ep(T+V)}$ to any order in $\ep$ via
 \be
\e^{\ep( T+ V)}=
\prod_{i=1}^N
\e^{t_i\ep T}\e^{v_i\ep V},
\label{prod} 
\ee
assuming that the effect of $\e^{\ep T}$ and $\e^{\ep V}$ 
can be computed exactly.
The set of coefficients $\{t_i,v_i\}$ are determined by
the order condition. For classical dynamics (\ref{tvop}), 
every factorizations of the
form (\ref{prod}) produces a symplectic 
integrator\cite{ruth83,neri87,cre89,fr90,yos90,wis91,yos93,hairer02,mcl02,lr04}
which is an ordered sequence of alternating displacements 
of $\bp$ and $\bq$ preserving Poincar\'e invariants.
For periodic motion, their energy errors are bounded
and periodic, in contrast to explicit Runge-Kutta type
algorithms whose energy error grows linearly
with the number of periods\cite{shita,gladman}.
However, for solving time-irreversible
evolution equations such as (\ref{imeq}) or (\ref{fpeq})
with $T=\nabla^2/2$, the Green's function 
\be
G(\br',\br;t_i\ep)
\propto {\rm e}^{-({\bf r}^\prime-{\bf r})^2/(2t_i\ep)}
\ee
is the
diffusion kernel only if $t_i$ is positive.
If $t_i$ were negative, then the kernel is unbound and 
there is no way of simulating the diffusion process backward in time.

Historically, symplectic integrators were developed
extensively for use in classical 
dynamics\cite{ruth83,neri87,cre89,fr90,yos90,wis91,yos93,hairer02,mcl02,lr04}.
Since classical dynamics is time-reversible, there was no
impetus for demanding that all $t_i$ be positive. Moreover,
Sheng\cite{sheng89} and Suzuki\cite{suzuki91} 
have proved that all factorizations of the form (\ref{prod}) 
beyond second order must contain
some negative coefficients in the set $\{t_i, v_i\}$.
Goldman and Kaper\cite{goldman96} later proved that for
factorizations of the form (\ref{prod}) beyond second order, 
{\it both} operators must have at least one 
negative coefficient. Thus all conventional splitting schemes
beyond second order must contain some negative coefficients and none
can be used to solve time-irreversible problems. Because of
the Sheng-Suzuki theorem, it is also difficult to see how one
can devise all positive coefficients, {\it forward} time-step
algorithms.

The operator product (\ref{prod}) has the general
Campbell-Baker-Hausdorff expansion, 
\be
\prod_{i=1}^N
\e^{t_i\ep T}\e^{v_i\ep V}
=\exp\ep\biggl( \e_T T+ \e_V V+\ep e_{TV}[T,V]
+\,\ep^2 e_{TTV}[T,[T,V]]+\ep^2 e_{VTV}[V,[T,V]]+\cdots\biggr)
\label{prodform} 
\ee
where all the error coefficients $e_T$, $e_{TV}$, $e_{VTV}$,
etc., are calculable functions of $\{t_i,v_i\}$, in particular,
\be
e_T=\sum_{i=1}^Nt_i\qquad{\rm and}\qquad e_V=\sum_{i=1}^Nv_i.
\ee
In order for the product to be consistent with the original
evolution operator, the coefficients $\{t_i,v_i\}$ must
satisfy the above constraints with $e_T=1$ and $e_V=1$.
Forcing the remaining error coefficients to vanish results in
order conditions that $\{t_i,v_i\}$ must satisfy. It is easy
to force $e_{TV}=0$. Any left-right symmetric product will do.
For example,
\be
{\cal T}_2(\ep)=
\e^{\frac12 \ep\, T}
\e^{\ep V}
\e^{\frac12 \ep\, T}=\e^{\ep\bigl(T+V
\,-\frac1{24}\ep^2[T,[T,V]]+\frac1{12}\ep^2[V,[V,T]]+\cdots \bigr) }
\la{secfac}
\ee
produces the following second order symplectic algorithm according to
(\ref{tvop}):
\begin{eqnarray}
{\bf q}_1&=&{\bf q}_0+{1\over 2}\ep\, \frac{{\bf p}_0}m \nonumber\\
{\bf p}_1&=&{\bf p}_0+\ep\,{\bf F}({\bf q}_1)\label{sec}\\
{\bf q}_2&=&{\bf q}_1+{1\over 2}\ep\, \frac{{\bf p}_1}m\nonumber
\end{eqnarray}
where the last numbered variables are the updated
variables.
Thus any symmetric splitting with $e_T=e_V=1$ will result in
at least a second order algorithm.
The surprise is that, as first shown by
Sheng\cite{sheng89}, beyond second order a general sum 
of products of the form (\ref{prodform}) is incompatible 
with having positive coefficients. More specifically, 
Suzuki\cite{suzuki91} shown that the two error coefficients $e_{TTV}$ 
and $e_{VTV}$ cannot both be forced to zero for positive coefficients
$\{t_i,v_i\}$. Since Takahashi and Imada\cite{ti84} have shown that 
$[V,[T,V]]=|\nabla V(r)|^2$ is a local potential function when solving the
imaginary time Schr\"odinger equation, Suzuki suggested\cite{suzu95} that 
this error commutator be kept and ways be found to eliminate
$[T,[T,V]]$.

Following up on Suzuki's suggestion, 
this author derived three simple fourth-order 
forward algorithms\cite{chin97} in 1997 and 
demonstrated their efficiency in solving Kepler's orbit. Interestingly,
it was found that classically $[V,[T,V]]$ produces a force 
(also with potential $|\nabla V(r)|^2$)  
first derived by Ruth\cite{ruth83} via canonical transformations.
The two forward schemes $A$ and $B$ derived 
in Ref.\cite{chin97} were also known to Suzuki\cite{suzu96} 
based on McLachan's result\cite{mcl952} on slightly perturbed 
Hamiltonians. However, Suzuki did not implement
them to do any calculation. 

Since 1997, fourth-order forward algorithms have been widely applied to 
time-irreversible systems such as the Fokker-Planck equation
in deriving the first fourth-order Langevin algorithm\cite{for01},
the Kramers equation\cite{for012} for describing stochastic dynamics,
the Diffusion Monte Carlo algorithm\cite{for012,lest05} for
solving quantum many-body ground states,
the grid based imaginary time Schr\"odinger\cite{chin013} equation
in doing density functional calculations,  
Path-Integral-Monte Carlo methods\cite{jang01,chin04,sak04,cuer05,bon06,mezz06}
for computing the quantum trace at finite temperature, short 
time evolved wave 
functions\cite{chincif03} for doing variational quantum many-body
calculations, the Gross-Pitaevskii equation for describing
a Bose-Einstein condensate in a rotating anisotropic trap\cite{chin057}
and electrons in a magnetic field confined by 
quantum dots\cite{chin056} and rings\cite{chin064}.
These forward fourth-order algorithms are far more accurate than
second order algorithms and allow very large step sizes
to be used.

While forward algorithms are indispensable for
solving time-irreversible equations, from their inception\cite{chin97}
they have also been shown to be efficient in solving
time-reversible equations. In comparison with 
explicit Runge-Kutta algorithms and conventional non-forward symplectic
integrators, forward algorithms have been shown to be 
superior in solving the Kepler problem\cite{chin97,chin2000, chin054},
gravitational few-body problems\cite{chin052}, the real time
Schr\"odinger equation\cite{chinc01}, the real
time Schr\"odinger equation in a laser field\cite{chinc02,bay03,gold04}
and specially the radial 
Schr\"odinger equation\cite{chin061}. The incorporation of the
commutator $[V,[T,V]]$ in forward algorithms has also inspired
a new class of higher order {\it gradient} symplectic
integrators\cite{chinc02,ome02,ome03,ome06}. These new algorithms,
through not forward beyond fourth-order, have less backward steps
at higher orders.

The reason why forward integrators are  
also better in solving time-reversible, classical dynamics 
problems is not well understood. From the perspective of 
the operator product approximation (\ref{prod}), as long as the 
second-order error terms are zero, any symmetric factorization scheme
will be fourth-order; it should not matters whether these error terms
are forced to zero with positive or negative coefficients. However, just
as in the discussion of time-reversible and time-irreversible 
algorithms, one must move beyond the purely algebraic discussion of 
factorization schemes to examine how the resulting algorithms
produce the solution of any particular equation. This work uncovers 
a crucial difference between forward and non-forward schemes when 
they are implemented as integrators for solving classical dynamics
problems. For classical dynamics, if the trajectory is the exact solution to 
(\ref{eqpair}), then the force is evaluated {\it only} along the 
trajectory. As will be shown below, non-forward algorithms, when 
compared to forward algorithms at the same finite step size $\ep$, 
evaluate the force at intermediate positions  
far from the trajectory. As one reduces the size of the 
negative time steps, one also reduces the distance 
of these force evaluation points from the trajectory. 
In all cases examined, non-forward algorithms are improved by 
simply reducing the size of their backward time steps, allowing 
the force to be evaluated more closely along the trajectory.
One can argue that these intermediate force evaluation points
are not the trajectory outputs of the integrator and 
their placements are not required to be on the trajectory.
That is correct. However, since the exact trajectory is determined
{\it only} by forces evaluated on the trajectory, any unnecessary force
evaluation off the trajectory is just ``wasteful". One then
must reduce the time step to bring the force evaluation
points closer to the trajectory. This is exactly what is observed
in the following comparisons. To achieve the same accuracy,
non-forward integrators must use smaller time steps. The question here
is not about correctness; it is about efficiency.
In the following sections we will compare in detail both types
of integrators with three, four, five and six force evaluations. 
  
\section {Integrators with three-force evaluations}

We will begin by comparing the efficiency of integrators 
in solving the 2D Kepler problem defined by the Hamiltonian  
\be
H=\frac12 \bp^2-\frac{\bf q}{q^2}\,.
\ee
(Henceforth, we will always normalize the Hamiltonian with
kinetic energy $\bp^2/2$ and $m=1$.) This problem is
an excellent benchmark because one can gauge an integrator's
performance not just by its energy error but also by
its orbital precession error\cite{gladman,chin2000,chin054}. The latter is a more
direct measure of the accuracy of the computed orbit.
 
There are basically three fourth-order integrators 
that required only three-force evaluations: 1)  
the non-forward Forest-Ruth (FR) integrator\cite{cre89,fr90,yos90}, 
\be
{\cal T}_{FR}(\ep)=	{\cal T}_2(a_1\ep)
{\cal T}_2(a_0\ep)
{\cal T}_2(a_1\ep)
\la{fr}
\ee
where 
\be a_1=\frac1{2-2^{1/3}}\approx 1.35,\quad 
a_0=-\frac{2^{1/3}}{2-2^{1/3}}\approx -1.70\, .
\ee
2) The forward integrator A\cite{chin97,suzu96}:
\begin{equation}
{\cal T}_{A}(\ep)\equiv
  {\rm e}^{ {1\over 6}\ep V}
  {\rm e}^{ {1\over 2}\ep T}
  {\rm e}^{ {2\over 3}\ep \widetilde V}
  {\rm e}^{ {1\over 2}\ep T}
  {\rm e}^{ {1\over 6}\ep V},
\label{foura}
\end{equation}
with $\widetilde V$ defined by
\be
\widetilde V=V+{1\over 48}\ep^2 [V,[T,V]]
\label{superv}
\ee
corresponding to an effective force
\be
\widetilde {\bf F}(\bq)={\bf F}(\bq)+{1\over 48}\ep^2\nbb(|{\bf F(\bq)}|^2).
\label{modfa}
\ee
Transcribing each operator in (\ref{foura})
yields the integrator
\begin{eqnarray}
{\bf p}_1&=&{\bf p}_0+{1\over 6}\ep\, {\bf F}({\bf q}_0)\nonumber\\
{\bf q}_1&=&{\bf q}_0+{1\over 2}\ep\, {\bf p}_1\nonumber\\
{\bf p}_2&=&{\bf p}_1+{2\over 3}\ep\,
             \widetilde {\bf F}({\bf q}_1)\label{sima}\\
{\bf q}_2&=&{\bf q}_1+{1\over 2}\ep\, {\bf p}_2\nonumber\\
{\bf p}_3&=&{\bf p}_2+{1\over 6}\ep\, {\bf F}({\bf q}_2),\nonumber
\end{eqnarray}
Starting with initial values $\bp_0$ and $\bq_0$, the updated
variables are $\bp={\bf p}_3$ and $\bq={\bf q}_2$. Algorithm A only
requires two evaluations of the force and one evaluation of the force
gradient. Recently, Omelyan\cite{ome06} has suggested that 
the effective force (\ref{modfa}) can be evaluated by extrapolation 
\be
\widetilde {\bf F}(\bq)
={\bf F}\Bigl(\bq+{1\over 24}\ep^2{\bf F}(\bq)\Bigr)+O(\ep^4).
\label{modfaex}
\ee
The resulting integrator remains angular momentum and phase-volume conserving and
is nearly indistinguishable from a fully symplectic integrator when solving the
Kepler problem. We will denote this use of 
an extrapolated effective force (\ref{modfaex}) as algorithm A$^\prime$. Algorithm 
A$^\prime$ only requires three force evaluations. 3) The  Runge-Kutta-Nystrom
(RKN) integrator\cite{bat99}. If one expresses $\bp={\bf p}_3$ and $\bq={\bf q}_2$ 
directly in terms of $\bp_0$ and $\bq_0$, then (\ref{sima}) reduces to
\ba
&&\bq=\bq_0+\ep\,\bp_0+\frac16\ep^2\left(\bff_0+2\bff_1\right)\nn\\
&&\bp=\bp_0+\frac16\ep\left(\bff_0+4\bff_1+\bff_2\right).
\la{rkn}
\ea
This is the form of
the RKN algorithm, with $\bff_0=\bff(\bq_0)$,
$\bff_2=\bff(\bq_2)$, but with $\bff_1=\widetilde\bff(\bq_1)$. The conventional
RKN algorithm is defined by $\bff_0=\bff(\bq_0)$, $\bff_1=\bff(\bq_1')$
and $\bff_2=\bff(\bq_2')$, where
\ba
&&\bq_1'=\bq_0+\frac\ep{2}\bp_0+\frac12(\frac\ep{2})^2\bff_0\nn\\
&&\bq_2'=\bq_0+\ep\bp_0+\frac12\ep^2\bff(\bq_1')
\la{rknf}
\ea
are the estimated midpoint and final position respectively.

\begin{figure}
	\vspace{0.5truein}
	\centerline{\includegraphics[width=0.8\linewidth]{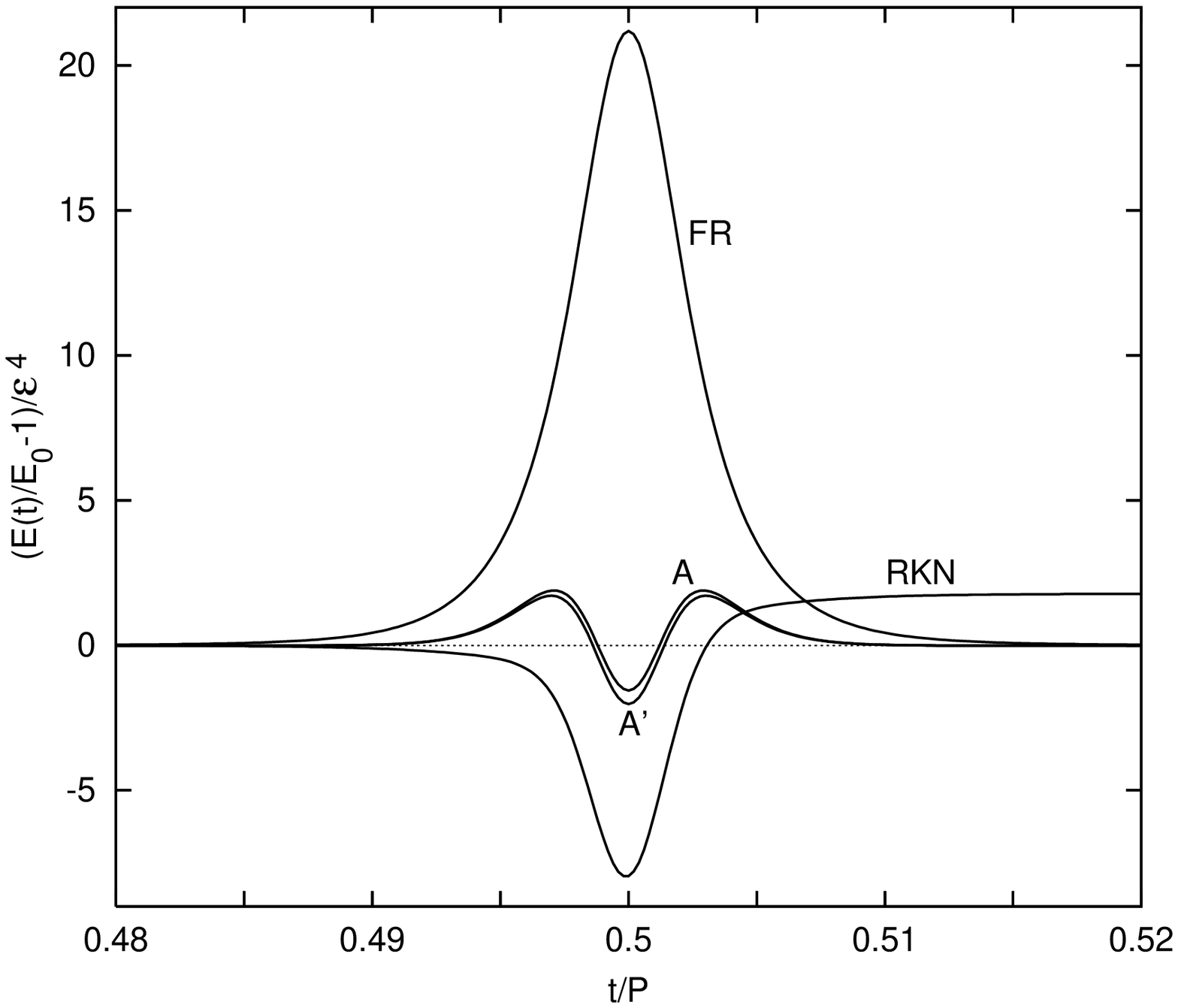}}
	\vspace{0.5truein}
\caption{The fourth-order energy error coefficients of forward
symplectic integrator A, extrapolated gradient algorithm
A$^\prime$, non-forward symplectic integrator FR (Forest-Ruth) and non-symplectic
integrator RKN (Runge-Kutta-Nystr\"om), as a function of time in terms of 
the orbital period $P$ when solving the 2D Kepler orbit. The time step size is
denoted by $\ep$. 
\label{eng3f}}
\end{figure}

Fig.\ref{eng3f} shows the fourth-order energy error coefficients of these four
algorithms at step size $\ep=P/5000$, where $P$ is the period 
of an highly eccentric orbit with initial values
$\bq=(10,0)$, $\bp=(0,1/10)$ and eccentricity $e=0.9$.
The error coefficient is obtained by dividing the energy
error by $\ep^4$ at smaller and smaller $\ep$ until a convergent
curve emerges independent of $\ep$. The curve is further normalized
by the initial energy. Thus each algorithm has a characteristic
error coefficient, its error ``fingerprint", in solving the 
Kepler orbit. For symplectic algorithms, this
convergence is already set in when $\ep\approx P/1000$. 
For the RKN algorithm, the energy error curve after the 
mid period keep on lowering with decreasing step size, showing
no sign of convergence at finite $\ep$. The error 
only spikes at mid period near the pericenter point. Non-symplectic
integrator such RKN are characterized by an irreversible
increase in the energy error after each period. The error spike
of the FR algorithm is nearly ten times as large as that
of algorithm A. The extrapolated gradient algorithm A$^\prime$ closely
matches that of A.

\begin{figure}
	\vspace{0.5truein}
	\centerline{\includegraphics[width=0.8\linewidth]{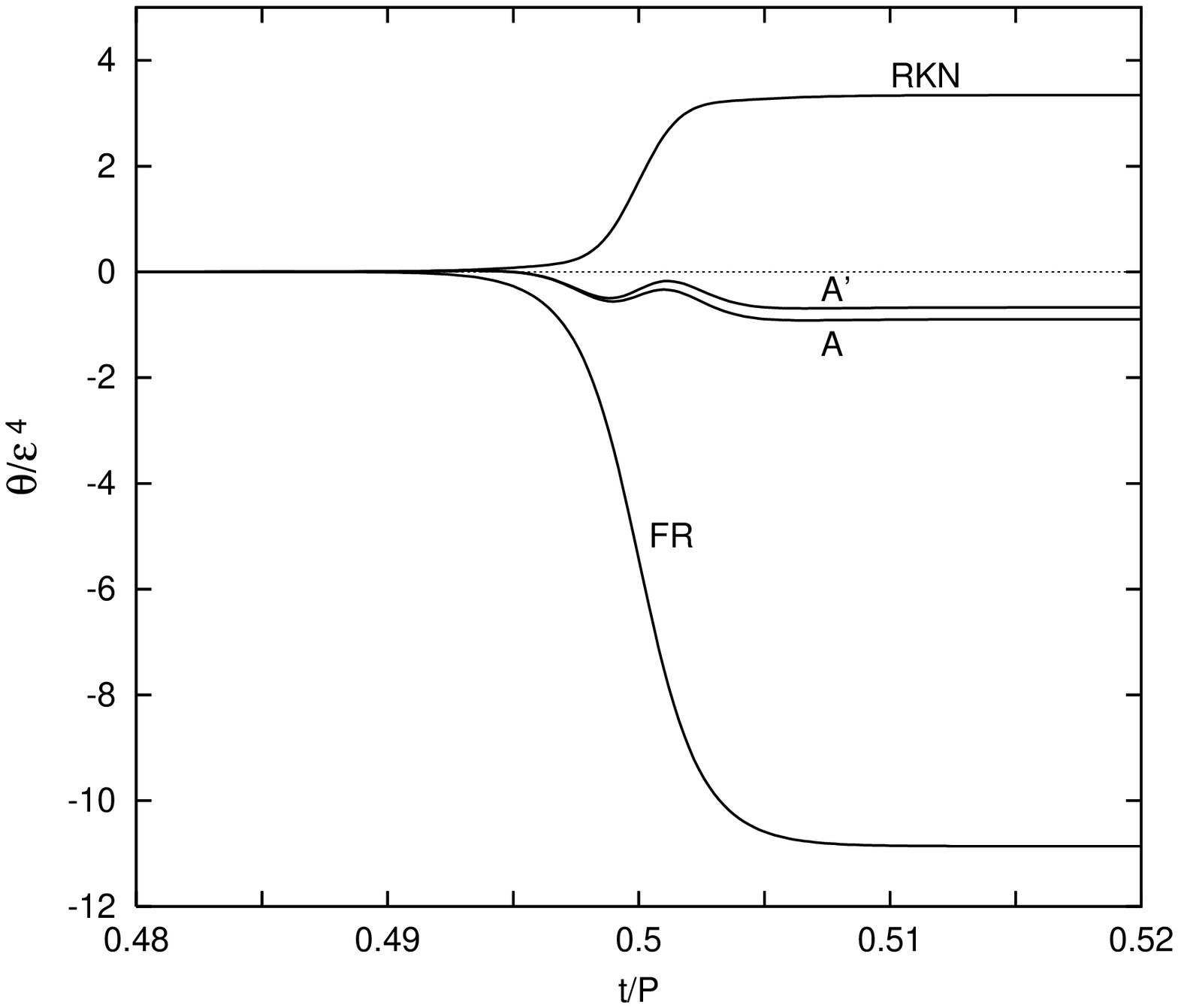}}
	\vspace{0.5truein}
\caption{The fourth-order orbital precession error coefficient
as measured by the rotation angle of the Laplace-Runge-Lenz (LRL) 
vector for integrators described in Fig.\ref{eng3f}. 
\label{ang3f}}
\end{figure}

Since all symplectic integrators have periodic energy errors,
the energy error is not the most critical benchmark.
The orbital precession error, as measured by the rotation of
Laplace-Runge-Lenz (LRL) vector\cite{chin2000,chin054,chin067}, 
is more discriminating. Fig.\ref{ang3f} shows
the rotation angle of the LRL vector along the trajectory of the
particle. The corresponding error coefficient is again extracted by
dividing by $\ep^4$. In this case, the RKN integrator shows 
the same convergence as the symplectic integrators. Thus the
precession error coefficient is well-defined
and irreversible for all algorithms. 
The FR integrator's error
is three times as large as that of RKN and 10 times as large as
integrator A and A$^\prime$. When the force gradient is extrapolated,
the energy error remains periodic, but 
the precession error can differs substantially. Here A$^\prime$'s
error is smaller, but in other cases it may not be. We will
revisit this point later.

\begin{figure}
	\vspace{0.5truein}
	\centerline{\includegraphics[width=0.8\linewidth]{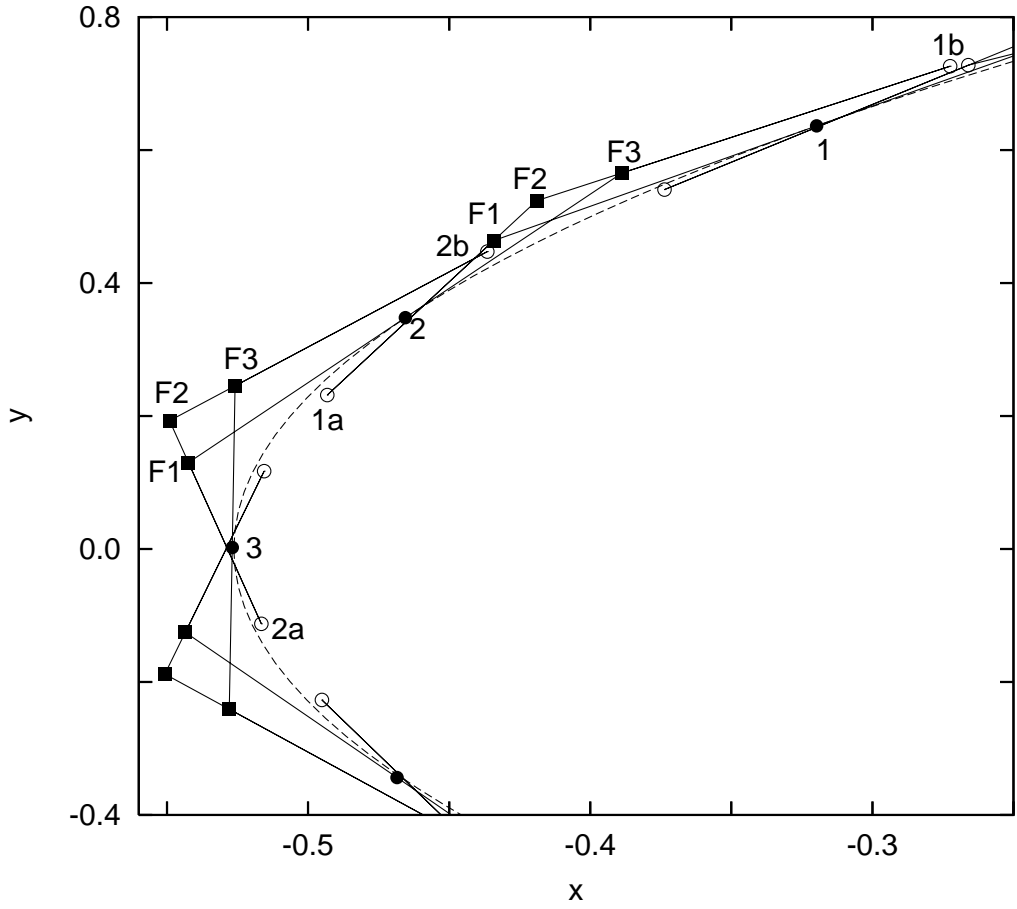}}
	\vspace{0.5truein}
\caption{The intermediate positions and force evaluation points of the
Forest-Ruth integrator. The dash curve is the exact orbit.
See text for details. 
\label{backfr} }
\end{figure}

\begin{figure}
	\vspace{0.5truein}
	\centerline{\includegraphics[width=0.8\linewidth]{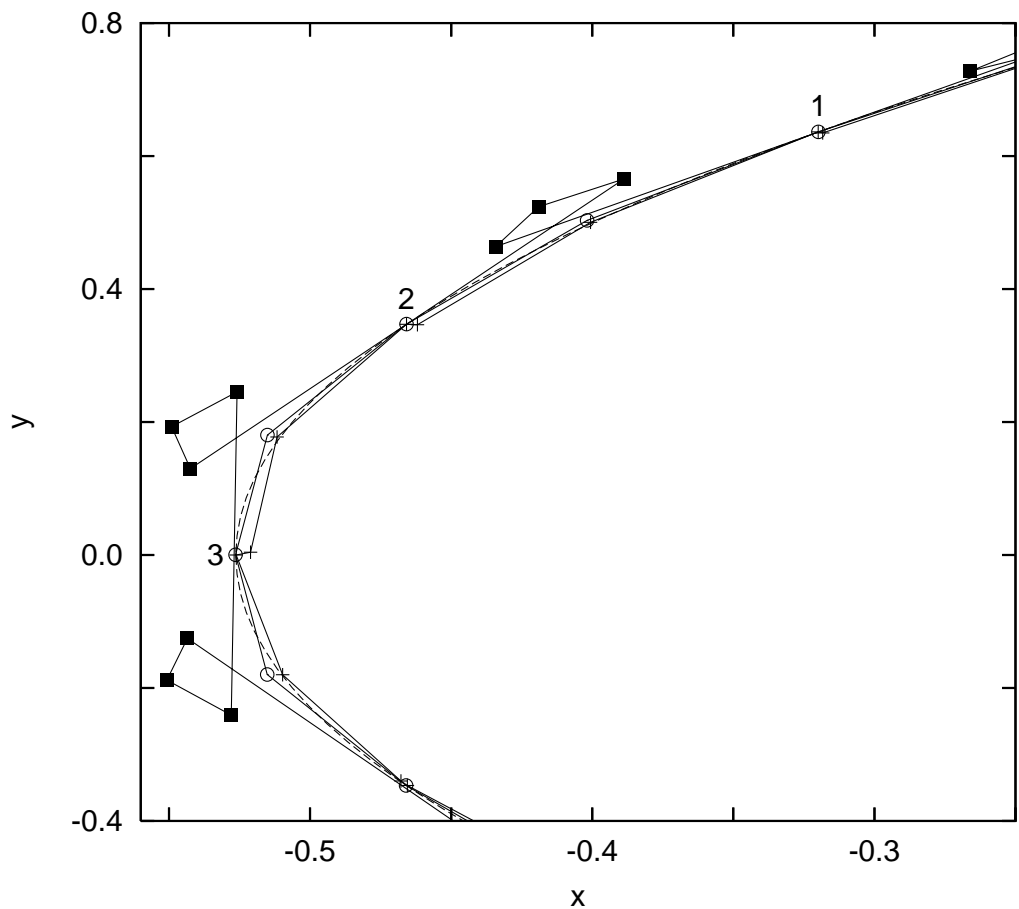}}
	\vspace{0.5truein}
\caption{The force evaluation points of integrator FR (solid squares),
 forward integrator A (circles) and non-symplectic integrator RKN (plus signs). 
\label{back3f}}
\end{figure}

In order to understand the poor performance of the FR integrator,
we track its approach toward the pericenter point 3 at
two time steps earlier at point 1, as shown in Fig.\ref{backfr}.
The time step used here is $\ep=P/400$.
The FR integrator consists of three applications of
${\cal T}_2(\ep)$, resulting in three
overlapping triangles. The first application of ${\cal T}_2(a_1\ep)$
begins at position 1, evaluates the force at F1, and lands at
1a. This is the triangle 1-F1-1a. The application of the backward 
substep ${\cal T}_2(a_0\ep)$ begins at 1a, evaluates the force
at F2 and brings trajectory back past the starting point to $1b$. 
This is the backward triangle 1a-F2-1b. The final application of
${\cal T}_2(a_1\ep)$ begins at 1b, evaluates the force at F3 and lands 
the trajectory at position 2. This is the final triangle 1b-F3-2. 
Starting at position 2, the algorithm repeats its three overlapping
triangles and zigzags its way to point 3. It is remarkable that 
FR can achieve fourth-order accuracy by such a tremendous zigzagging. 
Notice that as the FR algorithm tries to turn the ``corner" near 
the pericenter 3, all of its force evaluation points are {\it far off} 
the trajectory. In Fig.\ref{back3f}, the positions where each algorithm 
calculates the force are plotted. The backward loops executed 
by FR far off the trajectory is conspicuous. This is the 
fundamental reason by all non-forward algorithms perform poorly. 
By comparison, forward algorithm $A$ always evaluate the force and the 
force-gradient close to the exact trajectory. Since RKN is similar in form 
to $A$, it strays from the exact trajectory only near 
the pericenter point 3.

\section {Integrators with four-force evaluations}

\begin{figure}
	\vspace{0.5truein}
	\centerline{\includegraphics[width=0.8\linewidth]{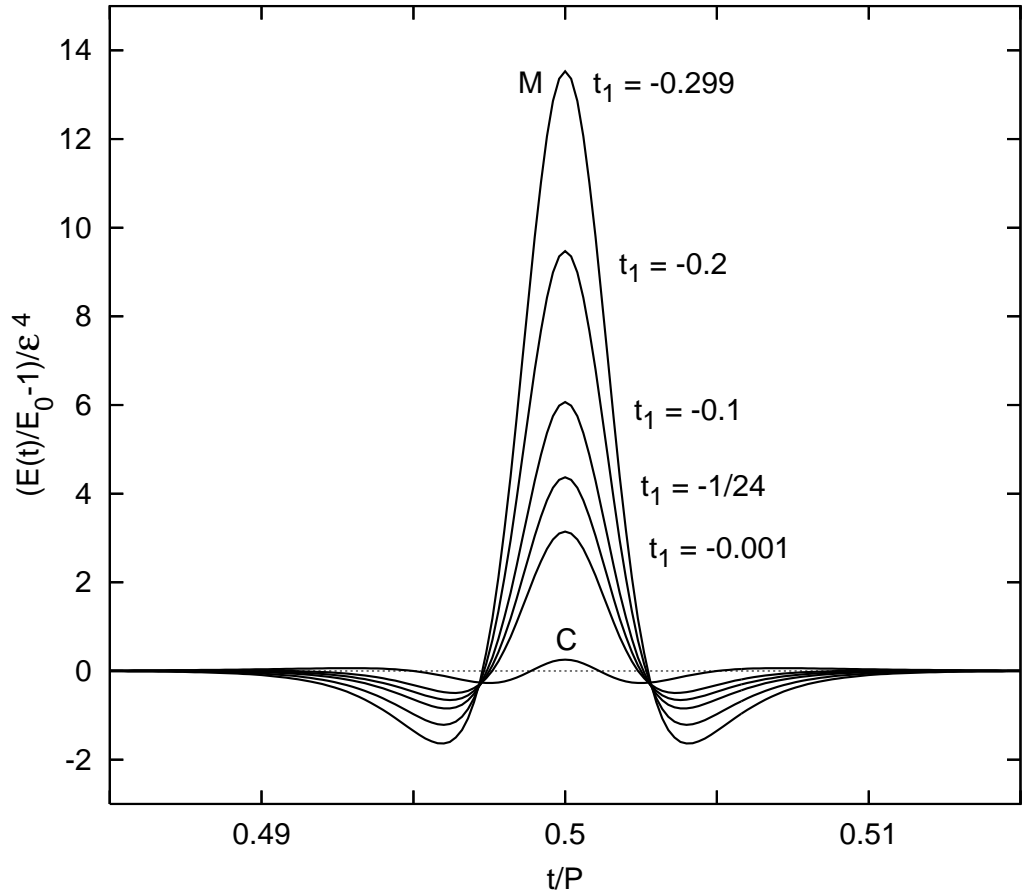}}
	\vspace{0.5truein}
\caption{The fourth-order energy error coefficients of a family of non-forward
integrators (\ref{algm1}) with four force-evaluations 
including that of McLachlan (M) as compared to that of forward integrator C. 
The parameter $t_1$ characterize the size of the backward time step for
updating the intermediate positions. 
\label{eng4f}}
\end{figure}

\begin{figure}
	\vspace{0.5truein}
	\centerline{\includegraphics[width=0.8\linewidth]{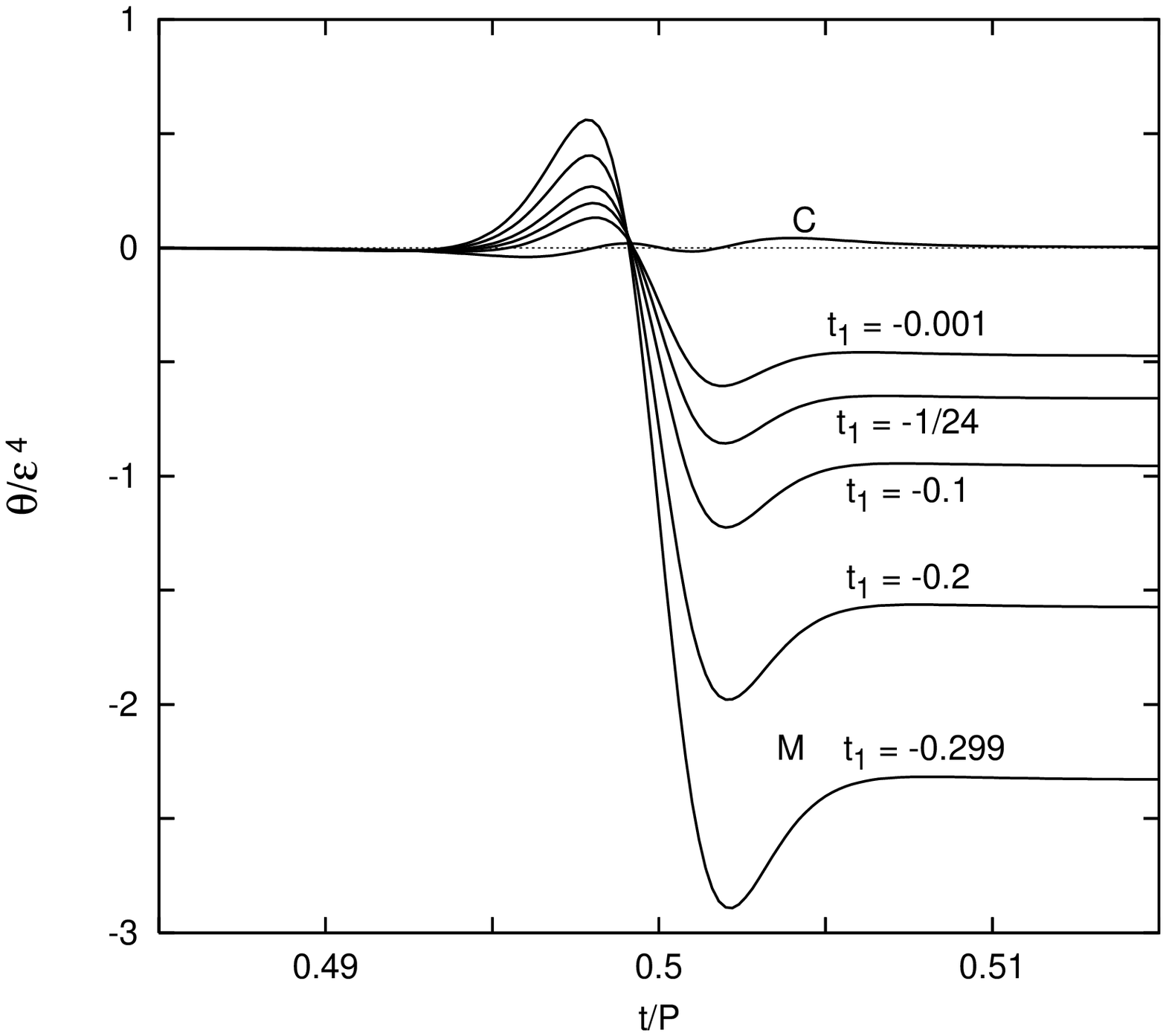}}
	\vspace{0.5truein}
\caption{
The fourth-order orbital precession error coefficient
as measured by the rotation angle of the Laplace-Runge-Lenz
vector for integrators described in Fig.\ref{eng4f}. 
\label{ang4f}}
\end{figure}

\begin{figure}
	\vspace{0.5truein}
	\centerline{\includegraphics[width=0.8\linewidth]{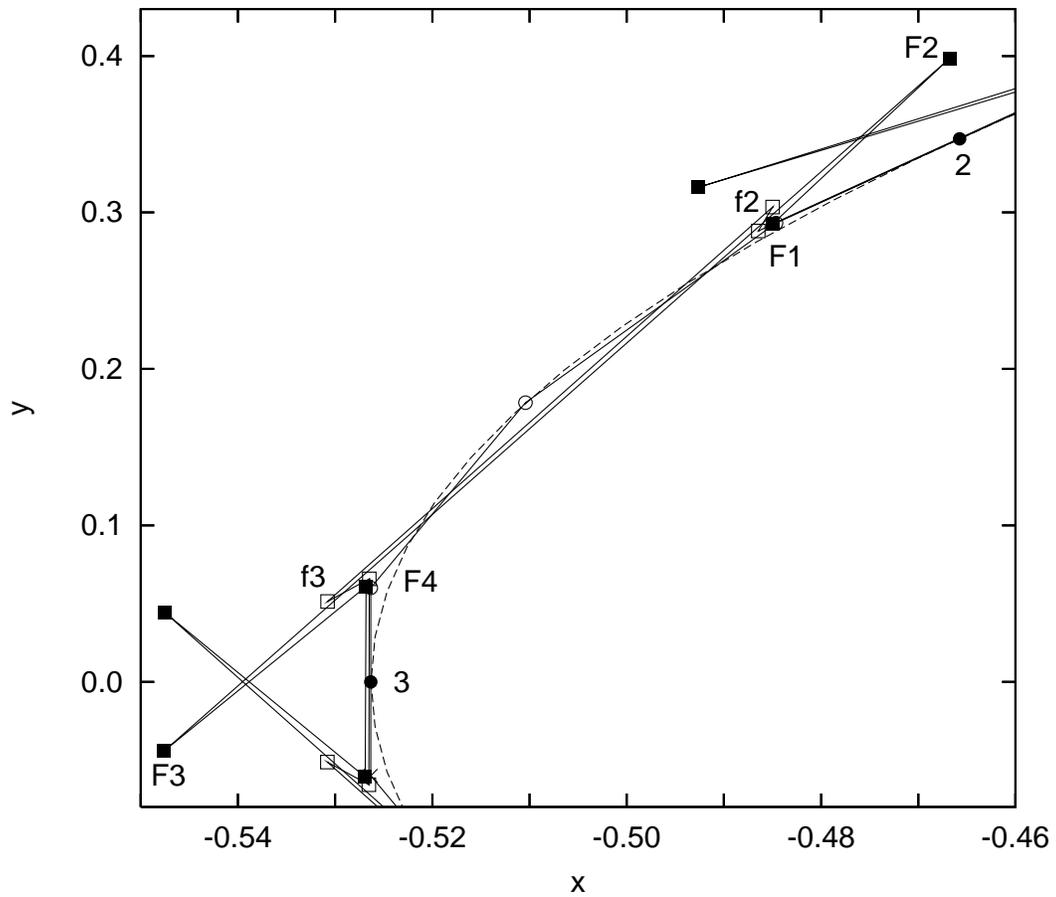}}
	\vspace{0.5truein}
\caption{The force evaluation points of integrator M (solid squares),
reduced backward time step integrator with $t_1=-1/24$ (hollow squares) and forward
algorithm C (circles). The solid circles denote the starting position 2 and
the final position 3 after one time step.
\label{back4f}}
\end{figure}

Because the error of the FR integrator is uncomfortably large, 
there is an ongoing effort to construct better non-forward 
algorithms by use of more force evaluations. A 
non-forward fourth-order
algorithm can be obtain by generalizing (\ref{fr}) to
\be
{\cal T}_4(\ep)=\prod_{i=1}^N{\cal T}_2(a_i\ep),
\la{gfr}
\ee
provided that the coefficients $a_i$ are left-right symmetric
satisfying\cite{cre89,yos90,suzu90} 
\be
\sum_{i=1}^N a_i=1 \quad{\rm and}\quad
\sum_{i=1}^N a_i^3=0.
\la{cubic}
\ee
Unfortunately, for $N=4$, there are no real solutions to the
above equations. We will examine algorithms of the general form
\be {\cal T}_{M1}= \dots 
\exp(\ep t_0 T) 
\exp(\ep v_1 V) 
\exp(\ep t_1 T) 
\exp(\ep v_2 V) 
\exp(\ep t_2 T), 
\la{algm1}
\ee
previously studied by McLachlan's\cite{mcl95}.
Since the algorithm is left-right symmetric, only operators
from the center to the right are indicated.	For a fourth-order
integrator, the order condition requires that
\be
v_1=\frac12-v_2,\quad
t_2=\frac16-4 t_1 v_1^2, \quad
t_0=1-2(t_1+t_2),
\ee
\be
w=\sqrt{3-12t_1+9 t_1^2},\quad
v_2=\frac14 \left(1\mp\sqrt{\frac{9 t_1-4 \pm 2w}{3t_1}}\,\,\right)
\ee
and that the free parameter $t_1<0$. There are four solution branches
for $v_2$. 
The choice of 
$$t_1=\frac{121}{3924}(12-\sqrt{471})\approx -0.299$$
with 
\be
v_2=\frac14 \left(1+\sqrt{\frac{9 t_1-4 + 2w}{3t_1}}\,\,\right)
\la{v2m1}
\ee
reproduces McLachlan's\cite{mcl95}
recommended algorithm. 
By simply reducing the size of the negative time
steps $t_1$, we obtain results as shown in Fig.\ref{eng4f} and Fig.\ref{ang4f}. 
The choice of $t_1=-1/24$ yielded simple
analytical coefficients 
\be
v_2=\frac{1+\sqrt5}4\quad{\rm and}\quad t_2=\frac{11-\sqrt5}{48}.
\ee
Also shown are results of forward
integrator C\cite{chin97}:
\be 
{\cal T}_{C}= \dots 
\exp({1\over 4}\ep \widetilde V)
\exp( {1\over 3}\ep T)
\exp( {3\over 8}\ep V)
\exp( {1\over 6}\ep T),
\la{chinc}
\ee
where $\widetilde V$ is as defined in (\ref{superv}) with the same
interpretation (\ref{modfa}). Algorithm C uses three force and
one force gradient evaluations. The force gradient can again
be extrapolated by another force evaluation. The results are similar 
and will be omitted in this comparison. To understand the
poor performance of non-forward algorithms, we again plot their
force evaluation points in Fig.\ref{back4f}. Starting at position 2, 
McLachlan's algorithm evaluates the force at F1, back tracks 
and evaluates the force the second time at F2, takes a giant
leap to F3, then back track again to F4. By reducing $t_1$ to
$-1/24$, the back tracking steps F2 and F3 are reduced to f2 and f3.
However, it is not possible to move any force evaluation
points closer to the midpoint of the trajectory. The force evaluation
points of algorithm C are indicated by circles. Its first and last
force evaluation points nearly coincide with F1 and F4, however, it
evaluates the force and the force gradient right at the midpoint of
the trajectory as shown by the unobstructed circle.

\begin{figure}
	\vspace{0.5truein}
	\centerline{\includegraphics[width=0.8\linewidth]{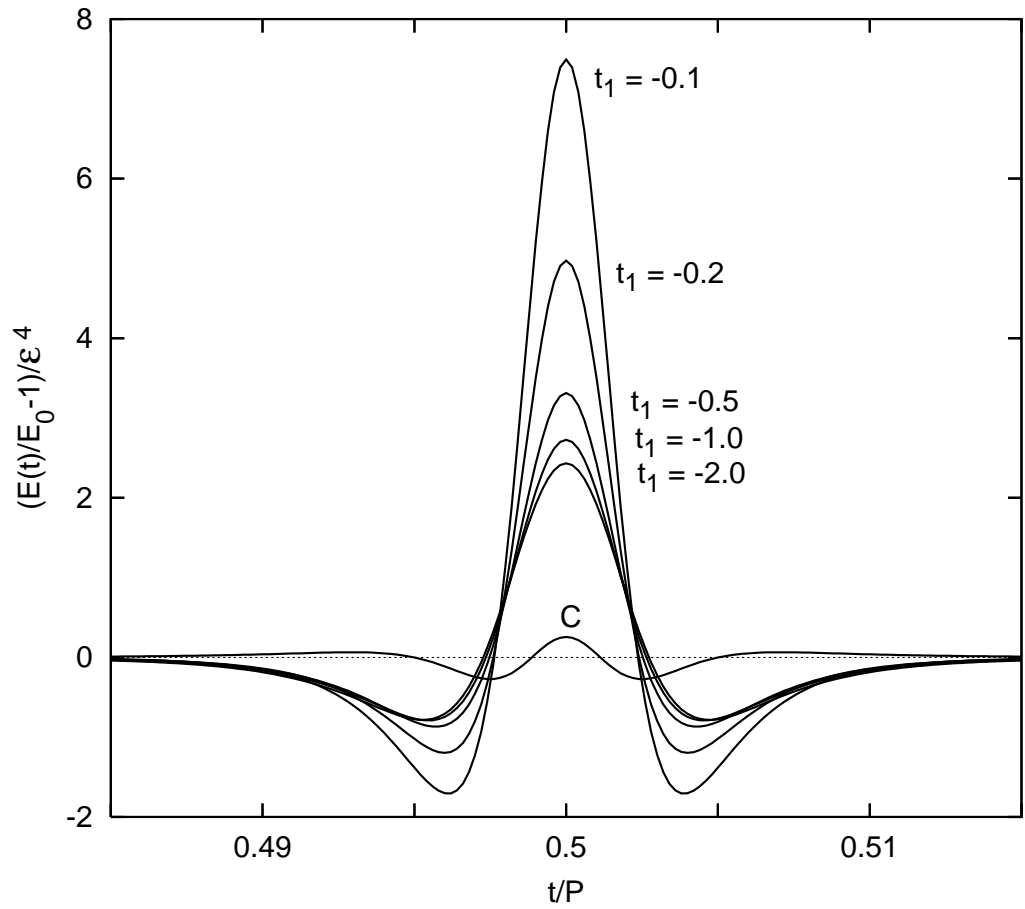}}
	\vspace{0.5truein}
\caption{The fourth-order energy error coefficients of a family of non-forward
integrators (\ref{algm2}) with four force evaluations which updates the 
{\it momentum} first. Here, the
more negative the parameter $t_1$ the smaller the negative time step size for 
updating the intermediate positions. 
\label{eng4f2}}
\end{figure}

\begin{figure}
	\vspace{0.5truein}
	\centerline{\includegraphics[width=0.8\linewidth]{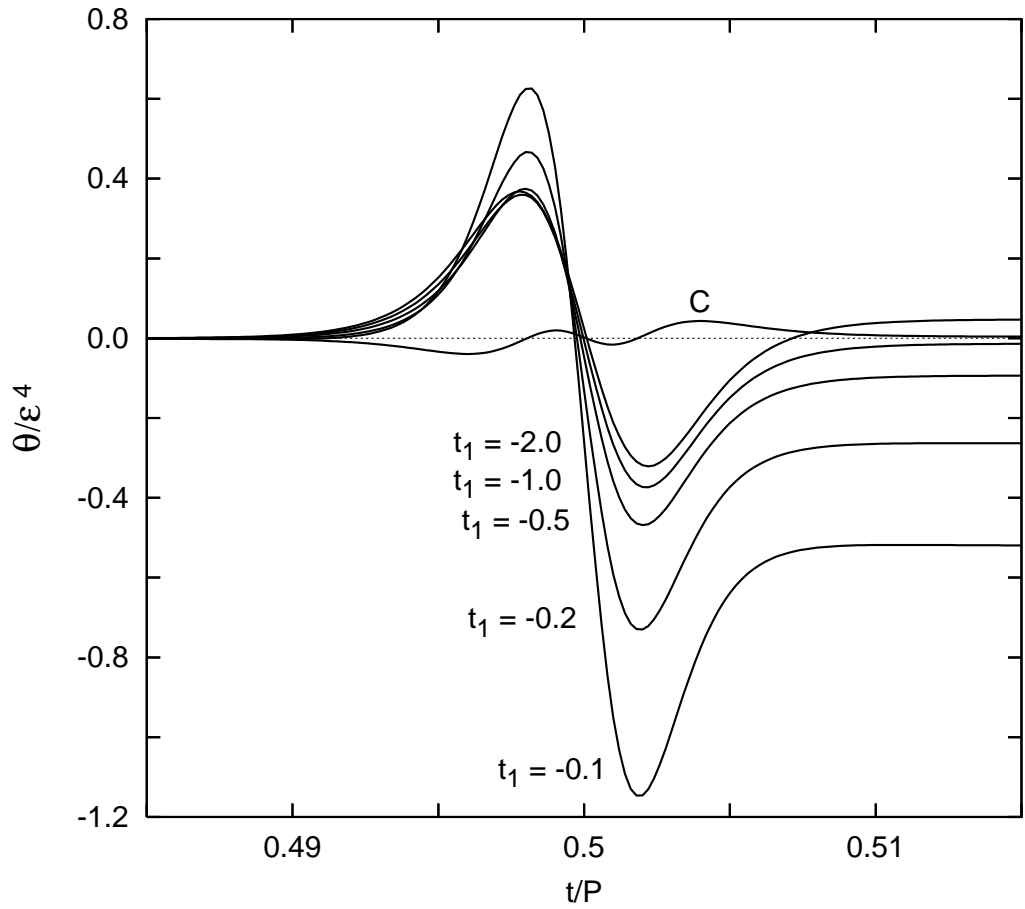}}
	\vspace{0.5truein}
\caption{The fourth-order orbital precession error coefficient
as measured by the rotation angle of the Laplace-Runge-Lenz
vector for integrators described in Fig.\ref{eng4f2}. 
In this case, it is possible
to fine tune $t_1$ so that the precession error is zero after each period.
\label{ang4f2}}
\end{figure}

\begin{figure}
	\vspace{0.5truein}
	\centerline{\includegraphics[width=0.8\linewidth]{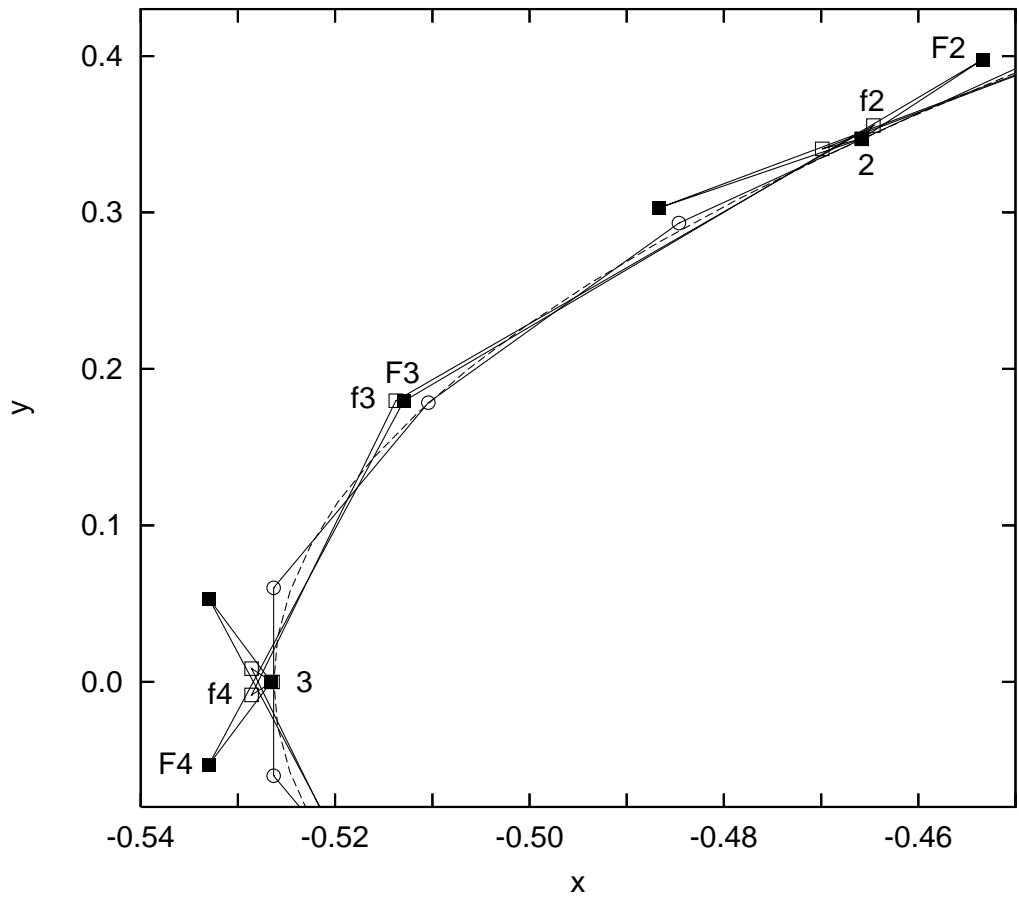}}
	\vspace{0.5truein}
\caption{The force evaluation points of integrator (\ref{algm2})
with $t_1=-0.1$ (solid squares), the
reduced backward time step integrator at $t_1=-0.5$ (hollow squares) 
and the forward algorithm C (circles).
\label{back4f2}}
\end{figure}

Better algorithms are obtained by interchanging $T\leftrightarrow V$
in (\ref{algm1}) 
\be {\cal T}_{M2}= \dots 
\exp(\ep t_0 V) 
\exp(\ep v_1 T) 
\exp(\ep t_1 V) 
\exp(\ep v_2 T) 
\exp(\ep t_2 V) 
\la{algm2}
\ee
so that the momentum is updated first with the choice
\be
v_2=\frac14 \left(1-\sqrt{\frac{9 t_1-4 + 2w}{3t_1}}\,\,\right).
\la{v2m2}
\ee
Now the force is evaluated initially, at two intermediate points,
and at the midpoint. The results, as shown in Fig.\ref{eng4f2} and 
Fig.\ref{ang4f2}, are much improved over the previous case. 
However, the locations of the forces evaluation points remain unusual. 
As shown in Fig.\ref{back4f2}, for 
$t_1=-0.1$, the algorithm first evaluates the force at the starting 
point 2, backtracks past 2 to evaluate the
force at F2, leaps forward to evaluate the force near the midpoint 
at F3, and shoots past the final point 3 to evaluate the force at F4.
Tuning the parameter $t_1$ more negative to $-0.5$ reduces the back 
tracking points from F2 to f2 and F4 to f4 and improves the algorithm.
However, as $t_1$ becomes even more negative, such as $t_1=-1$ or $-2$, those
back tracking points bunch up very close to the initial and final points 
and do not sample the trajectory evenly as algorithm C.

\section {Integrators with five and six force evaluations}

\begin{figure}
	\vspace{0.5truein}
	\centerline{\includegraphics[width=0.8\linewidth]{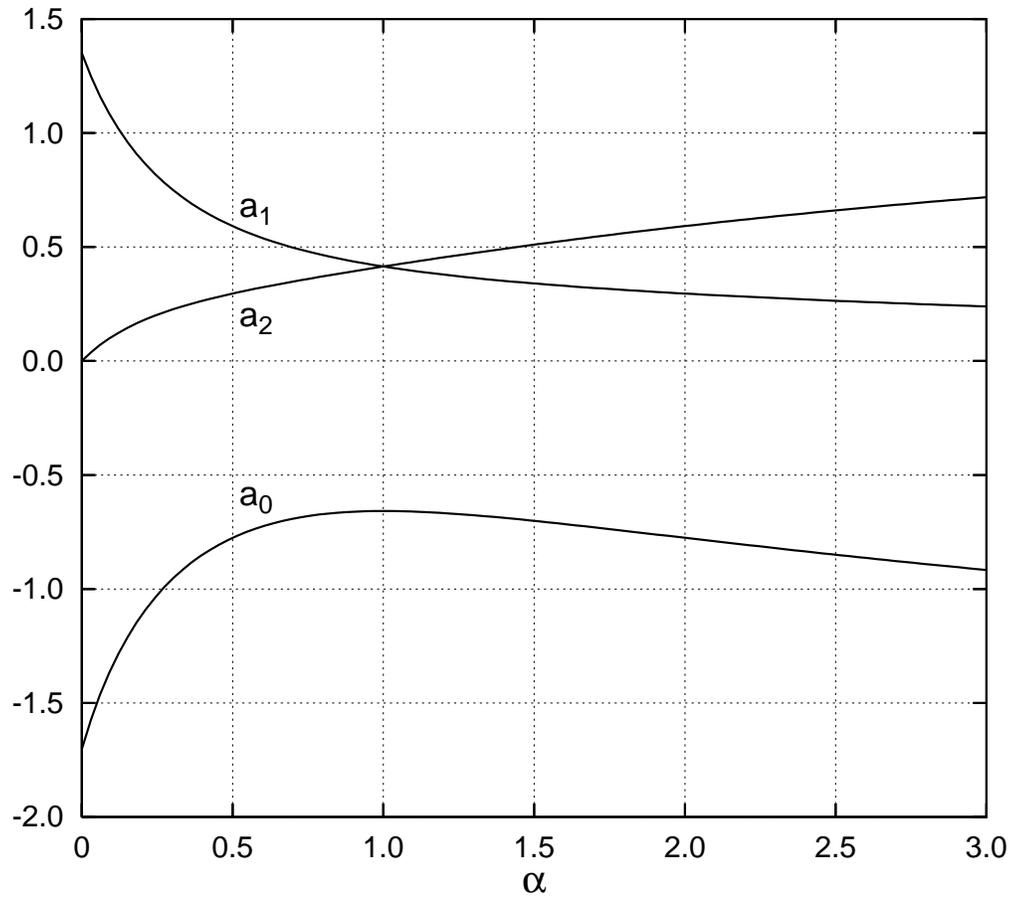}}
	\vspace{0.5truein}
\caption{The coefficients of integrator (\ref{gfr5}) as a function of
the free parameter $\alpha$. The negative coefficient $a_0$ is least
negative at $\alpha=1$.
\label{acoef}}
\end{figure}

\begin{figure}
	\vspace{0.5truein}
	\centerline{\includegraphics[width=0.8\linewidth]{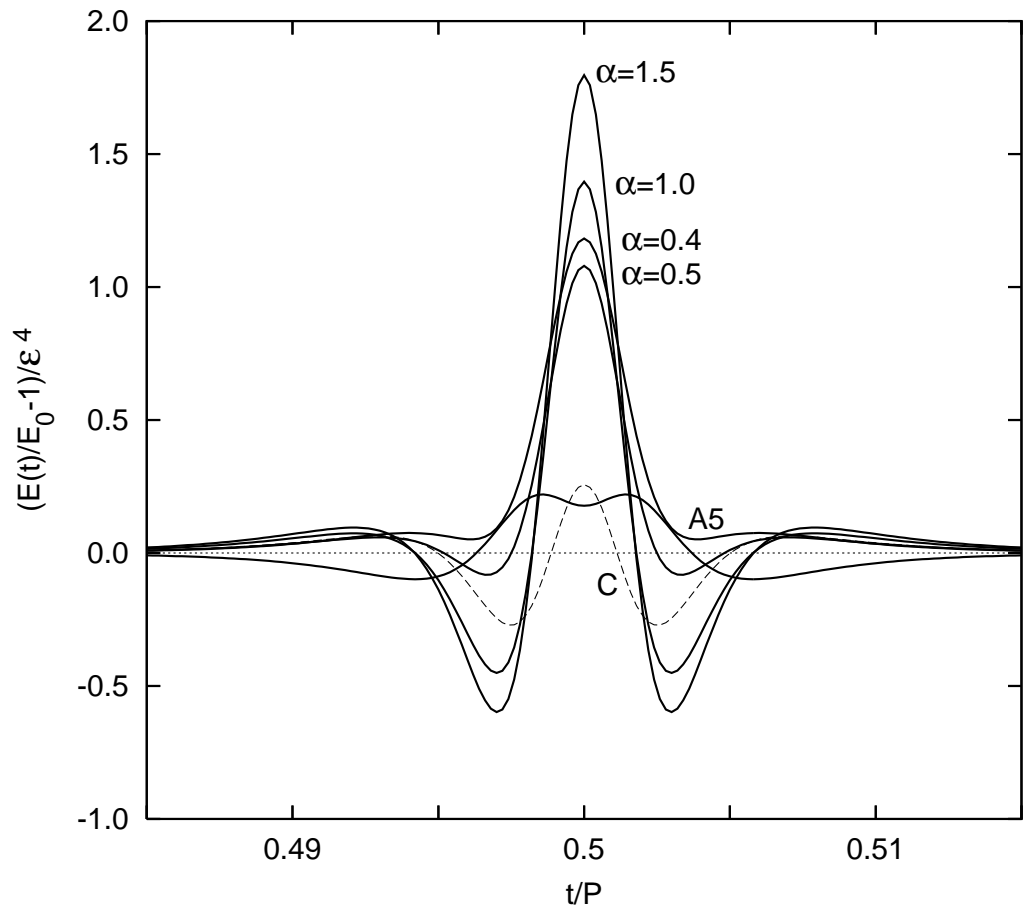}}
	\vspace{0.5truein}
\caption{The fourth-order energy error coefficients of a family of non-forward
integrators (\ref{gfr5}) with five force evaluations. These are compared to
the five-force forward integrator A5 and the four-force forward integrator C.
\label{eng5f}}
\end{figure}

\begin{figure}
	\vspace{0.5truein}
	\centerline{\includegraphics[width=0.8\linewidth]{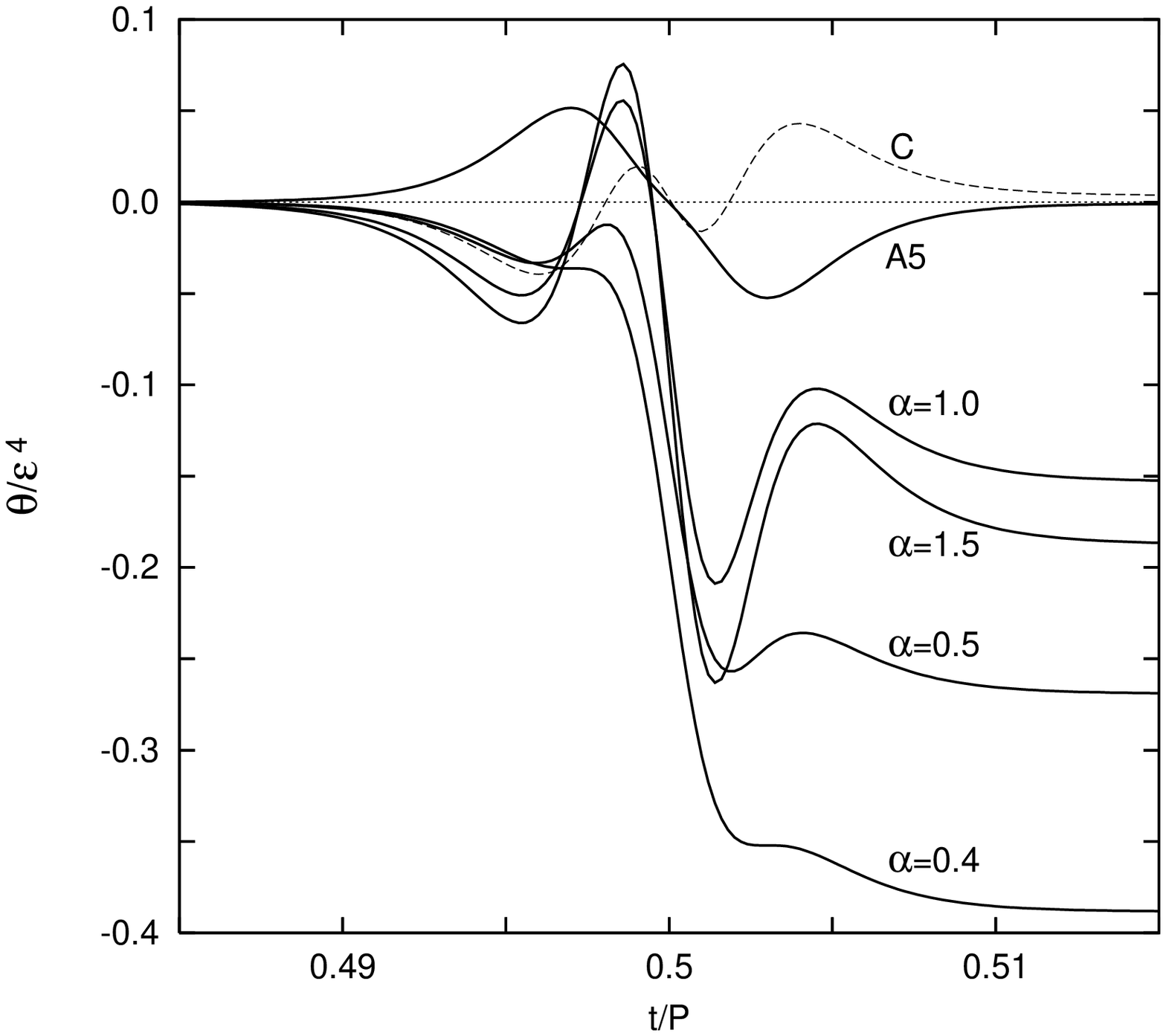}}
	\vspace{0.5truein}
\caption{The fourth-order orbital precession error coefficient
as measured by the rotation angle of the Laplace-Runge-Lenz
vector for integrators described in Fig.\ref{eng5f}. 
\label{ang5f}}
\end{figure}

For $N=5$, (\ref{gfr}) can be solved to give
\be
{\cal T}_4(\ep)=\dots {\cal T}_2(a_0\ep){\cal T}_2(a_1\ep){\cal T}_2(a_2\ep),
\la{gfr5}
\ee
with free parameter $\alpha$ and coefficients
\be
a_2=\alpha a_1,\quad
a_0=-2^{1/3}\left(1+ \alpha^3 \right)^{1/3} a_1,
\ee
\be
a_1=\frac1{2\left(1+\alpha\right)
-2^{1/3}\left( 1+ \alpha^3 \right)^{1/3}} .
\ee
The FR integrator is reproduced with $\alpha=0$. 
By introducing a non-vanishing 
$a_2$, one is able to reduce the negative step size $a_0$.
This is shown in Fig.\ref{acoef}. 
Fig.\ref{eng5f} and Fig.\ref{ang5f} show the energy and 
the precession error as a function of $\alpha$. While the
energy error height is lowest for $\alpha=0.5$, the procession error
is the smallest at $\alpha=1$. The latter is related to the fact that
the backward step size $a_0$ is minimized at $\alpha=1$. The
resulting algorithm with
\be
a_1=a_2=\frac1{4-4^{1/3}},\quad a_0=-\frac{4^{1/3}}{4-4^{1/3}},
\la{susumc}
\ee
has long been advocated by Creutz and Gocksch\cite{cre89},
Suzuki\cite{suzu90} and McLachlan\cite{mcl022}. 

\begin{figure}
	\vspace{0.5truein}
	\centerline{\includegraphics[width=0.8\linewidth]{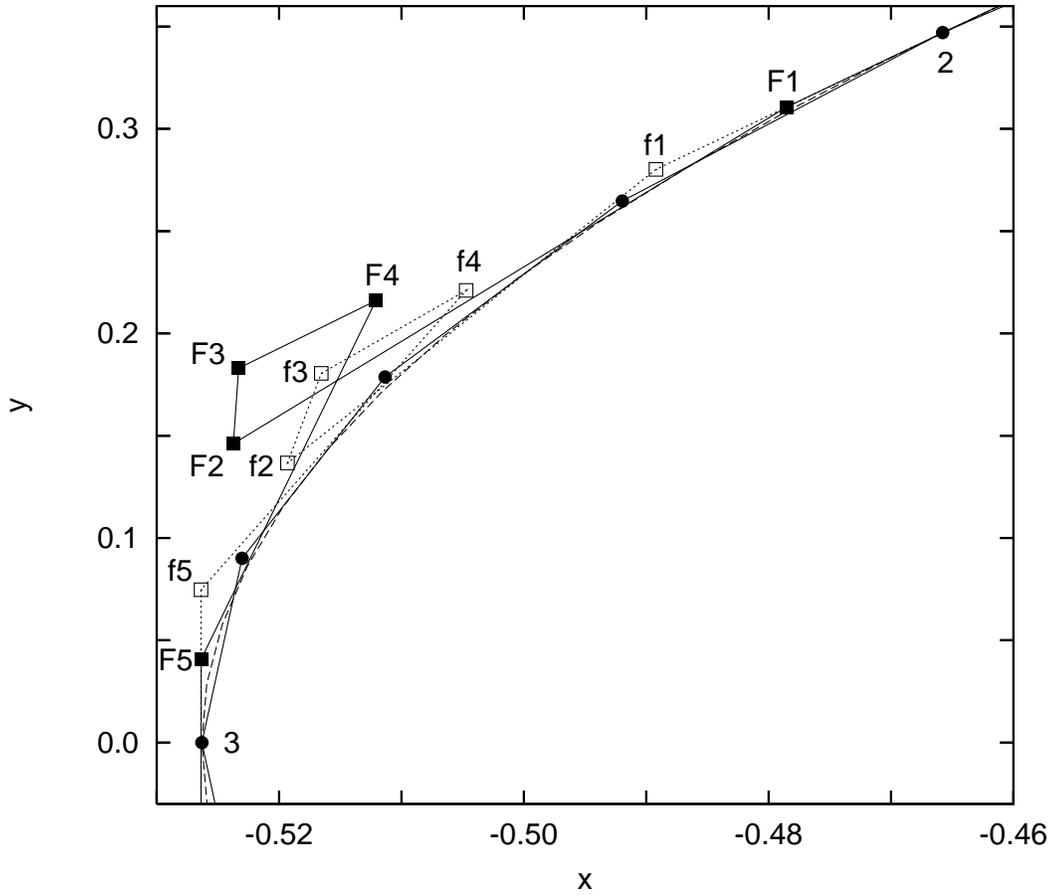}}
	\vspace{0.5truein}
\caption{The force evaluation points of 
non-forward integrator (\ref{gfr5}) 
at $\alpha=0.3$ (solid squares), with minimum
backward time step at $\alpha=1$ (hollow squares) 
and that of forward algorithm A5 (solid circles).
\label{back5f}}
\end{figure}

Recently, a fundamental theorem\cite{chin063} has allowed fourth order
forward algorithms to be derived for any number of operators\cite{chin062}.
In particular, one can generalize algorithm A to
$N-1$ force plus one force-gradient evaluations (or $N$ force evaluations
using extrapolation). This is the class of algorithm with uniform 
splitting coefficients
\be
t_i=\frac1{N-1},\quad v_i=\frac{N-1}{N(N-2)},
\ee
and where the algorithm begins and ends with a momentum
updating step:
\be
{\bf p}^\prime={\bf p}+\frac1{2N} \ep\left( {\bf F}({\bf q})
+{1\over 24(N-2)}\ep^2\nbb(|{\bf F(\bq)}|^2\right).
\ee
We will denote this class of algorithm as AN. The energy and precession
errors for A5 are as indicated in Fig.\ref{eng5f} and Fig.\ref{ang5f}. 
Algorithm A5's precession error is more than 4 times smaller than that of
algorithm C and 200 times smaller than non-forward algorithm
(\ref{gfr5}) at $\alpha=1$. As shown in Fig.\ref{back5f}, 
algorithms (\ref{gfr5}) again characteristically evaluates the force off 
the trajectory. As the negative time step $a_0$ is reduced by increasing 
$\alpha=0.3$ to $\alpha=1.0$, the off-trajectory force-evaluation 
triangle is reduced from F2-F3-F4 to f2-f3-f4.

\begin{figure}
	\vspace{0.5truein}
	\centerline{\includegraphics[width=0.8\linewidth]{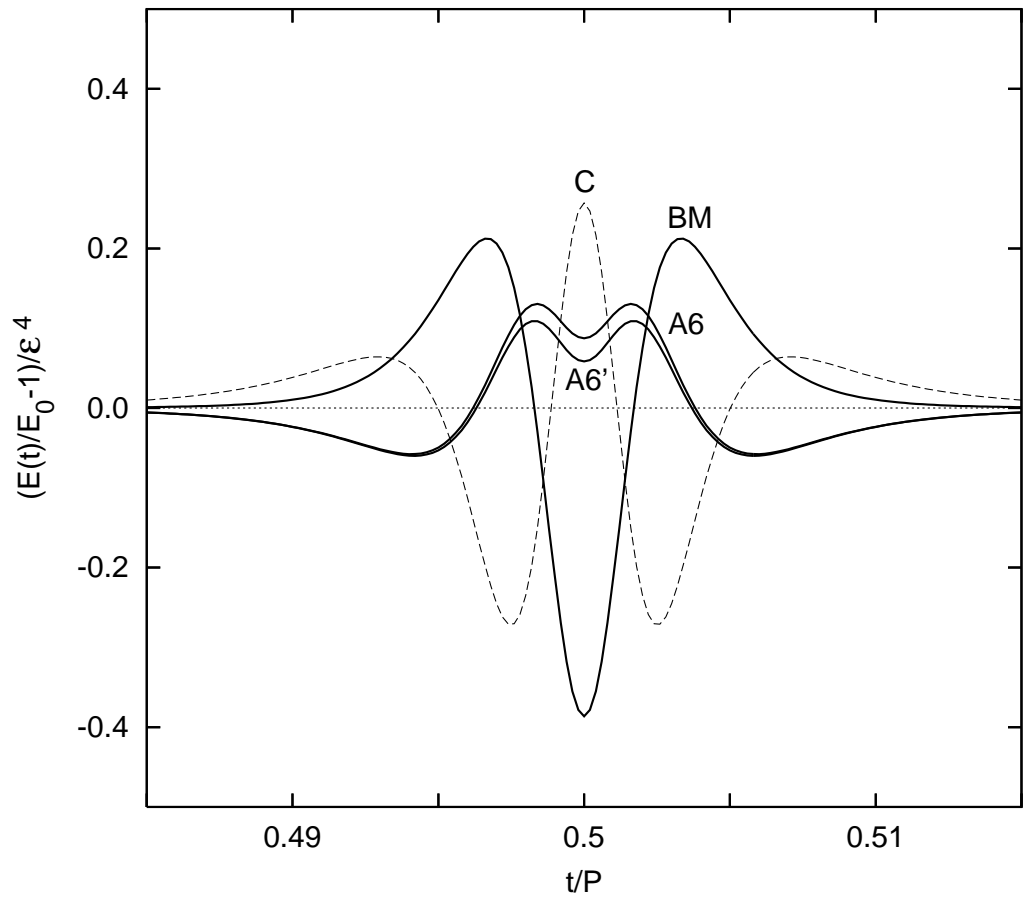}}
	\vspace{0.5truein}
\caption{The fourth-order energy error coefficients of 
three integrators with six force-evaluations.
BM is Blanes and Moan's integrator\cite{bm02}. A6 and A6$^\prime$ are 
forward integrators. A6$^\prime$ uses the extrapolated force gradient.
Algorithm C is a four-force forward integrator kept for comparison. 
\label{eng6f}}
\end{figure}

\begin{figure}
	\vspace{0.5truein}
	\centerline{\includegraphics[width=0.8\linewidth]{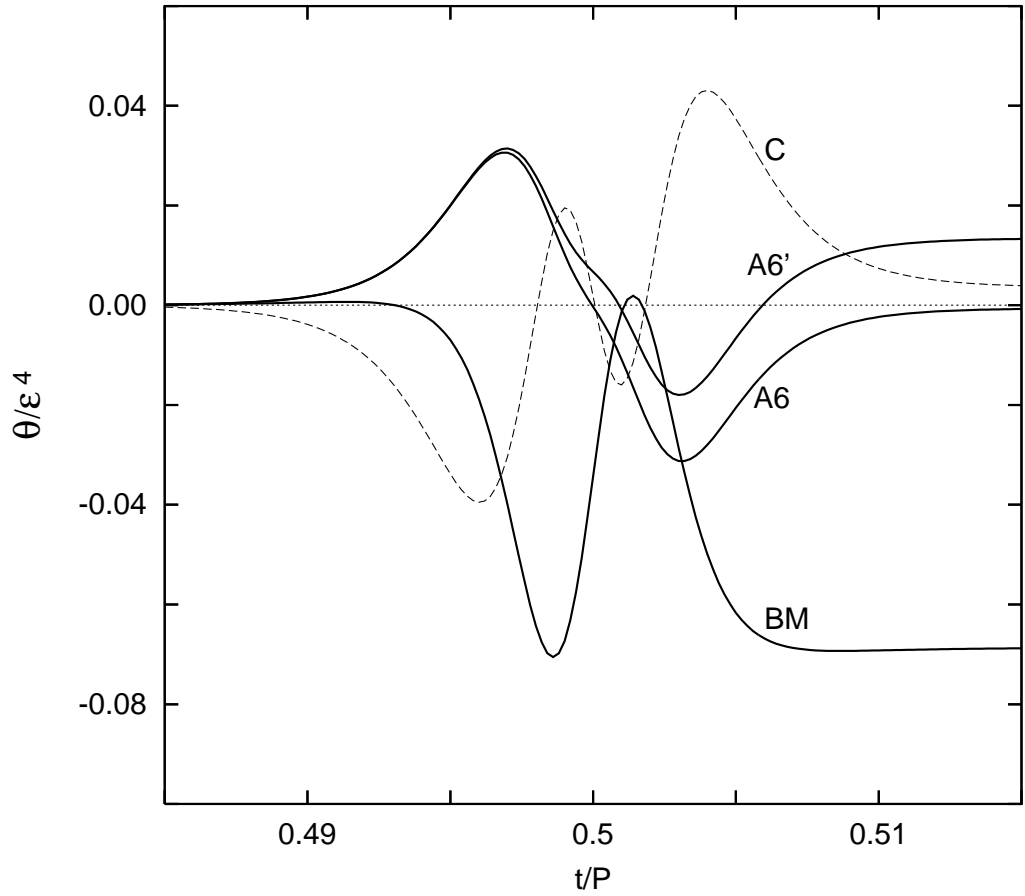}}
	\vspace{0.5truein}
\caption{The fourth-order orbital precession error coefficient
as measured by the rotation angle of the Laplace-Runge-Lenz
vector for integrators described in Fig.\ref{eng6f}.
The extrapolated gradient integrator A6$^\prime$'s precession error
is much larger than that of A6, but still smaller than that of BM. 
\label{ang6f}}
\end{figure}

\begin{figure}
	\vspace{0.5truein}
	\centerline{\includegraphics[width=0.8\linewidth]{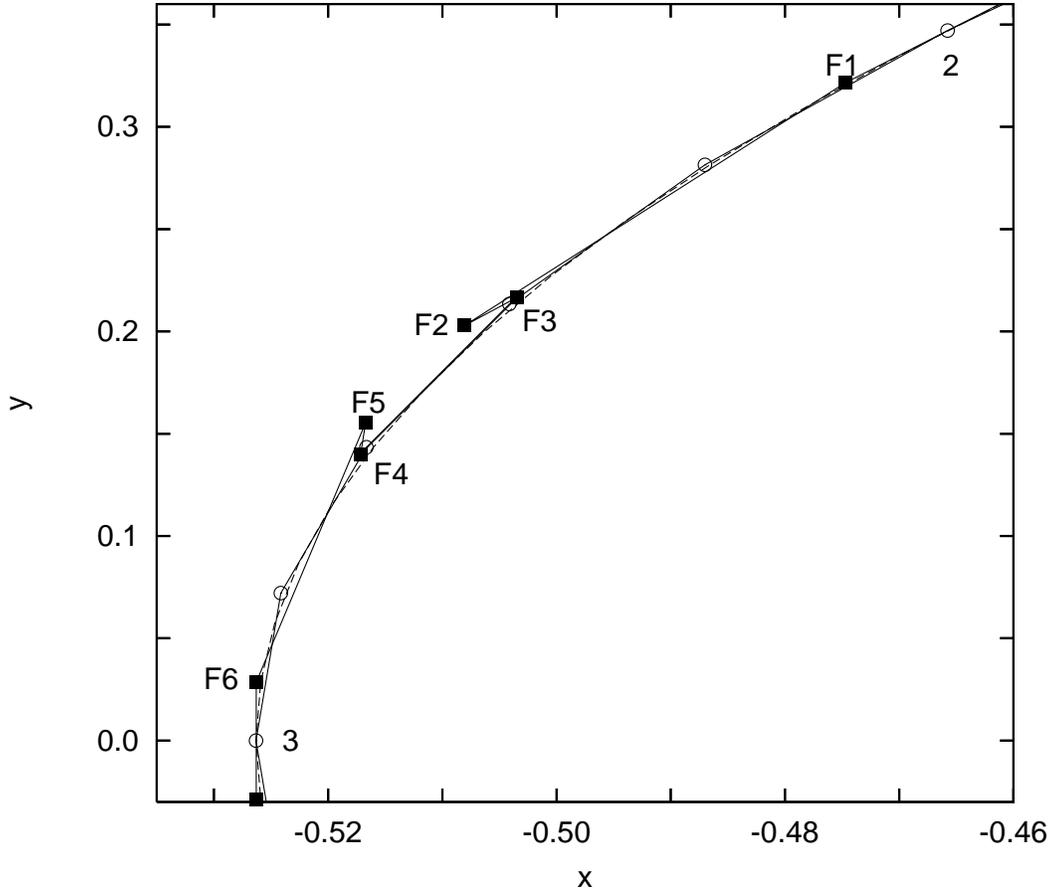}}
	\vspace{0.5truein}
\caption{The force evaluation points of 
non-forward integrator BM (solid squares) 
and that of forward integrator A6 (circles).
\label{back6f}}
\end{figure}

In Fig.\ref{eng6f} and Fig.\ref{ang6f}, we compare the energy and precession error
of A6 with that of Blanes and Moan\cite{bm02} (BM), a widely cited fourth-order
integrator with six force evaluations. BM's energy error is comparable to
that of algorithm C, but its maximum error height is four times that of
A6. For the precession error, BM's error is two orders of magnitude 
larger than A6 and twenty times larger than C. Included in the
comparison is algorithm A6$^\prime$, in which the force gradient 
is computed via extrapolation. Its energy error is nearly 
indistinguishable from that of A6, however, its precession error
is much larger. Blanes and Moan's integrator
is superior among non-forward algorithms because it has only
two very small backward time steps, as shown in Fig.\ref{back6f}.
(The momentum updating step first version of the BM integrator is not 
considered because it has much large errors than the position-first 
version discussed above.)  

\section {Conclusions}

All approximation methods for solving any evolution equation should
emulate its exact solution as much as possible. 
The efficiency of an algorithm cannot be decided on the
basis of factorization schemes, in which only the error coefficient
of the error commutators are known. In the past, 
symplectic integrators have been prized for their excellent 
conservation properties. However, because of the perceived
difficulty in circumventing the Shang-Suzuki theorem, forward
integrators were not developed until this decade. In this work,
we showed that forward integrators are more attuned to the exact
solution by evaluating the force closely on the trajectory. By
comparison, non-forward integrators, because of their backward 
time steps, are constrained to evaluate the force off the 
trajectory, resulting in the loss of efficiency. 
In all cases studied, non-forward integrators are improved by
simply reducing the size of their negative time steps.

\begin{acknowledgments}
This work is supported, in part, by a National Science Foundation 
grant, No. DMS-0310580.
\end{acknowledgments}

\bigskip
\bigskip
\centerline{REFERENCES}

\newpage
\newpage
\newpage
\newpage
\newpage

\newpage
\newpage

\newpage

\newpage
\newpage
\newpage
\newpage
\newpage
\newpage
\newpage
\newpage
\newpage
\end{document}